\documentclass[fleqn,10pt]{wlscirep}
\usepackage[utf8]{inputenc}
\usepackage[T1]{fontenc}
\title{Toward intelligent wireless networks in computer chassis}

\author[1,*]{Mohammadreza F. Imani}
\author[1,+]{Alexander L. Colson}
\author[1,+]{Leslie K. Miller}
\author[1,+]{Jorge A. Valdez}
\author[1,+]{Jose C. Sanchez}
\author[1,+]{Richard F. Rader}
\affil[1]{School of Electrical, Computer and Energy Engineering, Arizona State University, Tempe, AZ 85287,
USA.}
\affil[*]{mohammadreza.imani@asu.edu}
\affil[+]{These authors contributed equally to this work}

\begin{abstract}
Processing the exponentially growing amount of data produced daily requires efficient communication between different processing units in a computer. Traditionally, wired interconnects have been used to maintain these data links due to their energy efficiency and ability to support high data rates. However, as computing demands continue to increase in size and speed, these wired interconnects can become longer and less effective. One possible solution is to enhance the wired interconnects with short-range wireless communication (SRWC), which offers flexible resource allocation and the ability to broadcast data. However, implementing SRWC inside a computer chassis presents challenges due to multiple scattering. This scattering stretches the channel impulse response (CIR), leading to inter-symbol interference (ISI) and limiting data rates. To address this issue, we propose transforming the computer chassis into a smart radio environment by utilizing a reconfigurable intelligent surface (RIS). The RIS elements adjust the phase of reflected waves so that the multipath components combine at the receiver in a way that creates a pulse-like CIR. This approach has been experimentally validated within a typical computer chassis. The results of this study pave the way for integrating RIS-enabled SRWC to enhance wireless links in both current and future data processing units.

\end{abstract}
\begin{document}

\flushbottom
\maketitle
\thispagestyle{empty}

\section*{Introduction}

It is anticipated that we will produce the same amount of data that was produced in the last 100 years in less than ten seconds \cite{murray2018basic}. Processing this exponentially growing amount of data will require fast and efficient movement between different processing units such as multiple CPUs inside a computer chassis. The traditional solution to this demand, i.e., tungsten- or copper- based wires, are energy efficient and can support high data rate, but cannot scale up with the demand. Furthermore, wired interconnects are usually fixed and cannot reconfigure themselves, for example to allow for dynamic allocation of bandwidth to different links. As a result, any wired interconnect needs to be designed to support worst case scenarios, even if it may not be used in certain times. As computing demand, both in size and speed, grows exponentially, the wired interconnects become less efficient and longer---drawbacks that can result in latency. For example, wired interconnects require multiple hops for transferring data from one core to another. The limited data rate of wired interconnects thus may create a bottleneck that limits future computing systems. 

%Furthermore, wired interconnects cannot support dynamic allocation of resources for different links or allow for broadcast of data from one computing units to other

Recently, two possible solutions to augment the wired interconnects have been proposed: (i) free-space optical interconnects and (ii) short range wireless communication (SRWC) \cite{abadal2019wave,vlasov2012silicon,rommel2019beyond}. Free-space optical communication is especially attractive since it can allow for extremely fast data rates due to its ultra-wide bandwidth. However, optical interconnects require complex hardware and new material to be integrated into processing units. They are also usually large devices and suffer from thermal stability issues. Currently, optical interconnects are only used for links larger than 10 cm \cite{murray2018basic}. Furthermore, optical interconnects cannot broadcast data from one node to several nodes; they are usually useful for point-to-point communication and cannot be easily reconfigured \cite{abadal2019wave}. Use of SRWC as a possible solution for data transfer in computing units has gained traction due to its simplicity. Given the extensive effort put toward implementing future mobile networks such as 5G and beyond, SRWC enjoys a mature technology palette, allowing for implementing required hardware inside processing units. SRWC can thus offer communication links at different scales, from nodes on a chip to large scale data centers \cite{timoneda2018millimeter,narde2020antenna,narde2019intra}. Wireless communication can also support high data rate given their relatively large bandwidth. Furthermore, they are easier to integrate with conventional CMOS circuitry \cite{barakat2017innovative} and allow for broadcasting of data from one node to several nodes. They also allow for reconfigurable allocation of resources between links based on their usage. As a result, SRWC has great potential to facilitate fast and efficient data transfer in future processing units.

The primary issue facing wireless communication in data processing units is the presence of multiple scattering \cite{chiang2010short,timoneda2020engineer,f2022metasurface}.Multiple scattering occurs when the data processing unit, such as a chip package or computer chassis, behaves like a cavity for wireless signals. This multiple scattering within the enclosure results in stretched time-domain signals, leading to extensive delay spread. This can cause inter-symbol interference (ISI) and negatively affect the data transfer rate. While some techniques, such as orthogonal frequency-division multiplexing or time reversal, may alleviate some of the problems caused by multiple scattering \cite{alexandropoulos2021reconfigurable,bandara2023exploration,bandara2025trmac}, they can only be used for point-to-point communication links and are pair-specific. Furthermore, they require more complex modulation and demodulation, or waveform generation, increasing the cost and complexity of transceivers. As a result, a methodology to engineer the propagation environment in data processing units to reduce the impact of the multi-path on SRWC is highly desired.

In this paper, we aim to address the issues of SRWC for data processing units by transforming their propagation environment into a smart radio environment \cite{renzo2019smart,di2020smart,wu2021intelligent}. The underlying idea is to coat the environmental objects with reconfigurable metasurfaces---often referred to as reconfigurable intelligent surfaces (RISs)---to intelligently change the electromagnetic (EM) propagation environment. In essence, the objects and surfaces in the communication setting described above, which were considered limiting factors due to multiple scattering or blockages, can now be utilized to maintain or improve communication links. In this framework, the propagating environment, which has always been deemed out of touch, is now a design knob. This paradigm shift has sparked a whole new approach to engineering the wireless radio environment (i.e., smart radio environment). 

In multiple scattering environments such as those encountered by WiFi signals, these RIS reconfigure the effective reflection from its constitutive elements to change the multipath environment such that reflected waves interfere constructively and form a focus at the location of an intended user (i.e. over the air equalization) \cite{kaina2014shaping,dupre2015wave,del2016spatiotemporal,del2019optimally}. The ability of the RISs to overcome the impact of multipath and form a pulse-shape channel impulse response (CIR) has also been demonstrated in \cite{del2016spatiotemporal,alexandropoulos2021reconfigurable,f2022metasurface,faqiri2022physfad,tapie2024systematic}. In \cite{f2022metasurface}, it is shown that an RIS can form an intelligent wireless environment for intra-chip SRWC, while \cite{faqiri2022physfad} uses a physics-based model to show that RISs can form desired CIRs inside a room. In this paper, we expand the concept of an intelligent communication environment to include the computer chassis. Our experiments demonstrate that a Reconfigurable Intelligent Surface (RIS) placed inside a computer chassis can mitigate the effects of multiple scattering and create pulse-shaped Channel Impulse Responses (CIRs). This advancement opens up possibilities for integrating wireless communication within future computer chassis, supplementing traditional wired connections. %These results also experimentally confirm the operation predicted in [], [], paving the way for use of RIS-empowered SRWC for implementing high data rate in computing units of any scale.

\begin{figure}[ht]
\centering
\includegraphics[width=0.8\linewidth]{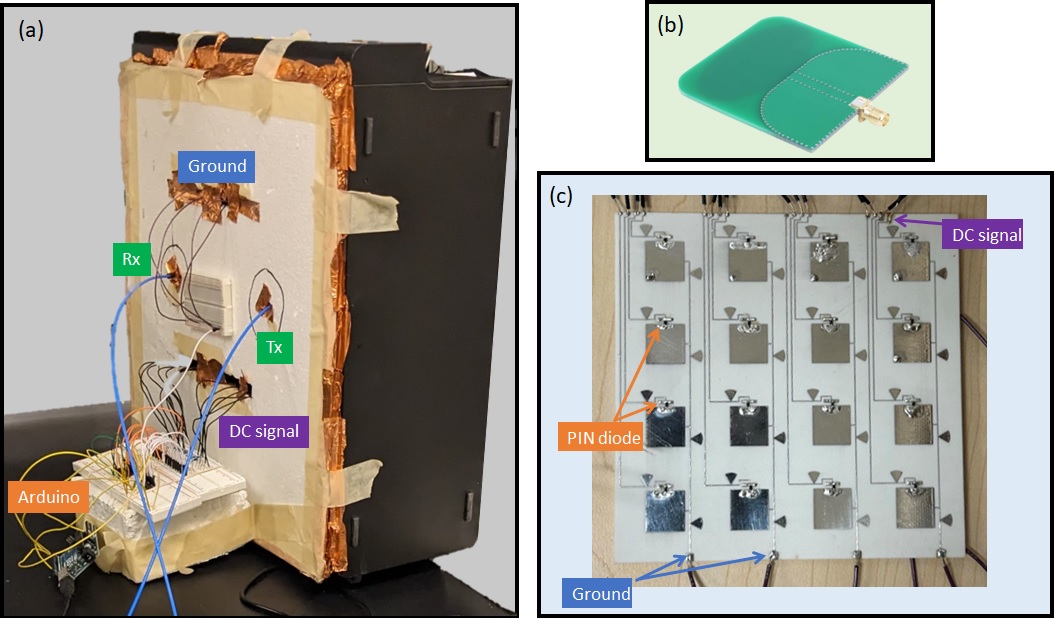}
\caption{(a) Measurement setup to examine the CIR inside a computer chassis. The white Styrofoam is covered in copper tape and is used to replace one side wall of the chassis. The RIS is placed near the center of the Styrofoam. (b) An ultra-wideband antenna is used to measure the channel response. (c) A picture of the fabricated RIS.}
\label{fig:general}
\end{figure}

\section*{Results}
The setup used to investigate SRWC inside a computer chassis is shown in Fig. \ref{fig:general} (a). In this setup, one sidewall of a computer chassis is replaced by a Styrofoam of the same size and covered in copper tape. Two antennas are placed at different random locations inside the computer chassis and act as a transmitter and receiver. For our demonstration purposes, we will connect these two antennas to two ports of a network analyzer to measure the channel between them. The antennas used in our demonstration are ultra-wideband ones developed by \href{https://www.elecbee.com/en-31285-UWB-Ultra-wideband-Antenna-Working-Frequency-2.4-10.5GHz?utm_term=&utm_campaign=shopping_%E7%BE%8E%E5%9B%BD2021/03/05&utm_source=adwords&utm_medium=ppc&hsa_acc=9958698819&hsa_cam=12473735731&hsa_grp=115457242501&hsa_ad=502747062194&hsa_src=g&hsa_tgt=pla-1186173532455&hsa_kw=&hsa_mt=&hsa_net=adwords&hsa_ver=3&gclid=CjwKCAiAjs2bBhACEiwALTBWZZLN3255AiAAsZr7efztauQMvoqMSGAtCHdsojc8S2xecgdEFj98dxoC-okQAvD_BwE}{Elecbee} and shown in Fig. \ref{fig:general} (b). The RIS used to engineer the propagation environment, discussed further in the Methods section, is shown in Fig. \ref{fig:general} (c). It consists of 16 patch elements, each loaded with a PIN diode. This RIS is attached to the Styrofoam and is addressed using an Arduino as shown in \ref{fig:general} (a). The Arduino DC signal can turn any of the diodes attached to the RISs elements \textit{on} and \textit{off}, giving rise to $2^{16}=65536$ distinct configurations of \textit{on} and \textit{off} elements. For brevity, we will refer to each arrangement of RIS diodes as a \textit{mask}. As discussed in the Methods section, the \textit{on} and \textit{off} states of the diodes toggle the effective response of each element between PEC and PMC (around 180 phase shift). This change in the boundary condition of the computer chassis (which is effectively a cavity) is the basis of operation of the proposed intelligent communication environment: as the effective boundary condition of the computer chassis changes, it also alters the channel between the two antennas. Our goal is to utilize this variation to optimize the response between the two antennas to mitigate multiple scattering and form a pulse-shaped channel impulse response (CIR).

\begin{figure}[ht]
\centering
\includegraphics[width=\linewidth]{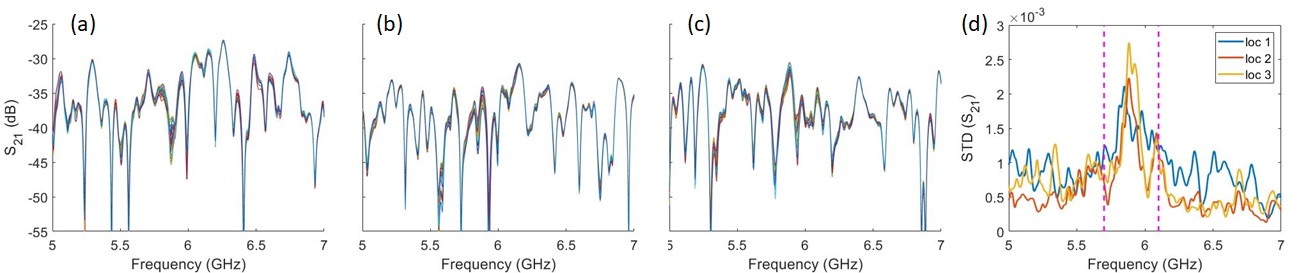}
\caption{(a)-(c) The frequency domain measurement between the transmitter and receiver antennas placed inside the computer chassis at different locations for 100 random masks. (d) The standard deviation of the measured response between the transmitter and receiver over 1500 masks.}
\label{fig:S21}
\end{figure}

First, we need to conduct an \textit{in situ} characterization of the RISs elements (as discussed in \cite{f2022metasurface}). The \textit{in situ}  characterization is necessary for multiple reasons. Firstly, due to PIN diode and fabrication tolerances, the resonance frequencies of the elements may have shifted. Furthermore, the setup used for the design of the elements described in the Methods section assumes an infinite array of elements which are excited by a linearly polarized normal plane wave (see Fig. \ref{fig:element}). This is different from the setup used in Fig. \ref{fig:general} (a) in which the RIS is placed inside an enclosure where the incident wave on the elements of the RIS can be from any arbitrary direction and polarization. Furthermore, each element of the RIS may have a different response, violating the infinite periodic assumption in the design setup. Lastly, the RIS is a finite $4\times 4$ array, as opposed to an infinite array assumed in the simulation. For the \textit{in situ} characterization of the RISs response, we examine the signal transfer between transmitter and receiver antennas in the setup shown in Fig.\ref{fig:general} (a) for the first $1500$ \textit{masks}. The result is shown in Fig. \ref{fig:S21} (a)-(c) for different locations of the transmitter and receiver inside the computer chassis. We can observe that the transmitted signal exhibits many nulls, consistent with the prediction from basic physics that the computer chassis will act similarly to a disordered electrically large cavity that can support many modes. We also observe that the signal between the two antennas changes as a function of the masks. These changes are unique to each antenna pair location but are primarily constrained between $5.7$ to $6.1$ GHz. This observation is better visualized when we examine the standard deviation of the transfer signal between the two antennas as shown in Fig. \ref{fig:S21} (d). Examining the result in Fig. \ref{fig:S21} (d), we can confirm that the proposed RIS's impact on the communication channel is more pronounced in the 5.7 to 6.1 GHz range. We thus select this frequency range for our investigation.

Having characterized the RIS, we now examine its utility in improving the CIR between the two antennas. To obtain the CIR, we perform an inverse Fourier transform of the channel’s
frequency measurement between 5.7 GHz to 6.1 GHz. To quantify the likeness of the CIR to a pulse, we define a figure of merit, FOM, as the ratio of the power in the main peak of the CIR to the power in the whole CIR. This FOM, which is also used successfully in previous works \cite{f2022metasurface,faqiri2022physfad}, can be described mathematically as 

\begin{equation}
    FOM=\frac{\int_{t_o-\Delta t/2}^{t_o+\Delta t/2} CIR(t)^2dt}{\int_{0}^{\infty} CIR(t)^2dt}
    \label{eq1}
\end{equation}
where $t_o$ is the time instance of the main peak and $\Delta t$ is a window defined around the main peak. We set the window empirically to be $\Delta t=0.286$ ns. It is worth noting that since we are interested in CIRs that resemble pulses, a small window is desired. Furthermore, we cannot calculate the CIR for infinite time. As a result, we only calculate the integral in the numerator of the equation above for $t=50 ns$, which was also set empirically based on the decay of the CIR observed in experimental results.

In previous works, optimization processes were introduced to obtain the RIS's mask that maximizes the FOM. This multi-step algorithm was especially needed since examining all possible masks was practically impossible (due to the large simulation time to model a scattering environment accurately). Here, we take advantage of the fact that we can examine all possible RIS configurations experimentally in a relatively much shorter time. In this manner, we ensure we have reached the global maximum. Since computer chassis are fixed environments, such an optimization process needs to be conducted once in practical implementation. 

\begin{figure}[ht]
\centering
\includegraphics[width=\linewidth]{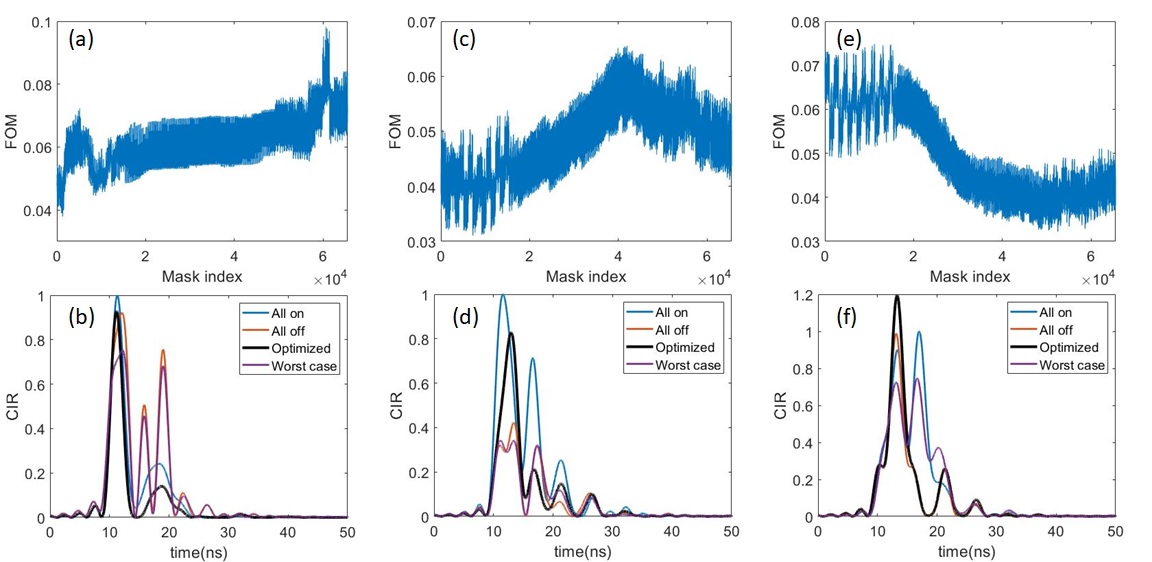}
\caption{(a), (c), and (e) Variation of FOM defined in (\ref{eq1}) as the RIS masks change for different locations of antenna pairs. (b), (d), and (f) are the corresponding CIR for example states of RIS for different locations of antennas pairs.}
\label{fig:CIR}
\end{figure}

The FOM and the corresponding CIR are shown in Fig. \ref{fig:CIR} for three different locations of transmitter and receiver antennas. We clearly see that the FOM changes as we alter the RIS's mask. The masks corresponding to the maximum or minimum FOM are different for each location of the antennas, which is consistent with the fact that we are dealing with a cavity environment. Furthermore, we observe that the FOM exhibits many local maxima that are close to the global maxima. This is also consistent with the fact that the multiple scattering environment exhibits mesoscopic correlation \cite{del2016spatiotemporal}. In Fig. \ref{fig:CIR} (b), (d), and (f), we have plotted the CIR for maximum and minimum FOM as well as for the cases when the diodes of the RIS are all \textit{on} or all \textit{off}. The latter two represent the cases when RIS approximates a PEC or PMC boundary condition. We clearly see that the calculated CIR is stretched over time and exhibits multiple peaks, some comparable to the main peak. Such a CIR will result in ISI and reduce the data rate. By using the proper RIS mask, we can mitigate the impact of multiple scattering and form a pulsed-shaped CIR that can be used for SRWC. This proposal is validated for three different locations of the antenna pairs inside the computer chassis as shown in Fig. \ref{fig:CIR}.

\section*{Discussion}

In this paper, we demonstrate the feasibility of transforming a computing unit into a smart radio environment. We have shown that by utilizing a simple RIS configuration with minimal power consumption, we can shape the desired CIR's characteristics. The results presented here also validate the predictions made using a physics-based model and full-wave simulations \cite{f2022metasurface,faqiri2022physfad}. With this experimental verification, we have established a novel approach to implement SRWC in data processing units. For the proof-of-concept presentation in this paper, the center frequency was selected to be around 6 GHz (for simplicity). In a practical implementation, a higher center frequency would be more appropriate. It is worth noting that extending the proposed RIS-empowered channel to higher frequencies is possible by redesigning the elements' geometry and selecting suitable switchable components. In fact, several recent works have demonstrated RISs that can operate at millimeter wave and THZ frequencies \cite{gros2021reconfigurable,sengupta2018terahertz}. However, we need to emphasize an important distinction here: the role of the proposed RIS is distinct from previous works in which the RISs are designed to form directive beams. The only requirement for the RIS elements is to exhibit tunable resonance near the resonance frequency. Satisfying this requirement is much easier than those required for beamforming RISs, especially at higher frequencies. We can thus anticipate a straightforward transition of this concept to higher frequencies. It is important to note that another requirement for previous RISs (which usually contributed to their cost and complexity) is their electrically large size in order to form directive beams with narrow beamwidth. In the proposed work, the RIS can be of any size, as long as it provides sufficient degrees of freedom (number of elements) to alter the propagation environment. If needed, one can add a few small RISs placed at different locations. As a result, the proposed RIS can be implemented at a much lower cost and complexity compared to previous works. 

While this work used a narrowband RIS with binary tuning, a more practical implementation can consist of elements loaded with varactor diodes to increase bandwidth, reduce ohmic losses, increase degrees of freedom, and decrease overall power consumption. Alternatively, one can use PIN diodes attached to elements with different resonance frequencies. Another limitation of the proof of concept unit discussed is its low degrees of freedom. One can easily increase the degrees of freedom in sculpting desired links by using RISs with more elements or by including multiple RISs. Another way to increase degrees of freedom is to use elements that engage with both polarizations. When the number of elements (or degrees of freedom increases, the brute force optimization process presented in this paper may not be suitable. Instead, one can use an algorithm similar to the one discussed in \cite{f2022metasurface,faqiri2022physfad}, where, after a few iterations among elements of the RIS, it arrives at a local optimum. The experimental results presented here do, in fact, show that many local optimums exhibit FOM similar to the optimum one. It is worth noting that such algorithms can be implemented using a very small number of measurements, allowing for the RIS-empowered network to be adaptable to changes inside the chassis (for example, if a cable is moved or removed).

Another future direction is the optimization of communication links for data broadcast inside computer chassis and servers (i.e., single input to multiple output or SIMO networks). This concept has been validated in WNOC \cite{tapie2024systematic,imani2021demand} and can be explored for chassis in the future. It is worth noting that the proposed concept can offer benefits other than improving SRWC data rate. The proposed device can also be used to detect intruders or pests inside processing units. Any such intruders would change the channel, which can be sensed by monitoring the transfer signal between the two antennas \cite{del2018dynamic}. Another possibility is to augment the computational power by conducting wave-based computation using the proposed RIS. For example, the proposed RIS can be used to implement differentiation, reservoir computing, or discrete Fourier transform on the stream \cite{del2018leveraging,ma2022short}.

An alternative application of the proposed RIS-empowered chassis is to overcome eavesdropping attempts on the computer processors. In fact, an emerging threat in recent years is the use of covert communication links to obtain data from air-gapped computer chassis. In these works, the spurious radiation from the CPU (for example, DRAM clock) is used to transfer information out of the computer chassis \cite{zhan2020bitjabber,camurati2022noise}. The RIS configuration proposed and demonstrated in this work can also be used to randomly alter the propagation environment, in order to reduce the data rate of covert communication links (which are usually assumed to have a deterministic channel). This RIS can be small and installed on one side of the computer chassis. Its operation is independent of the computer (and thus cannot be infiltrated) and consumes a very small amount of power. Since it does not radiate any signal, it does not interfere with the computer's operation. Investigating these interesting scenarios is left for future work. %Preliminary results have shown great promise [] and can be explored further in future.

\section*{Methods}
\subsection*{Element Design} 
\begin{figure}[ht]
\centering
\includegraphics[width=0.6\linewidth]{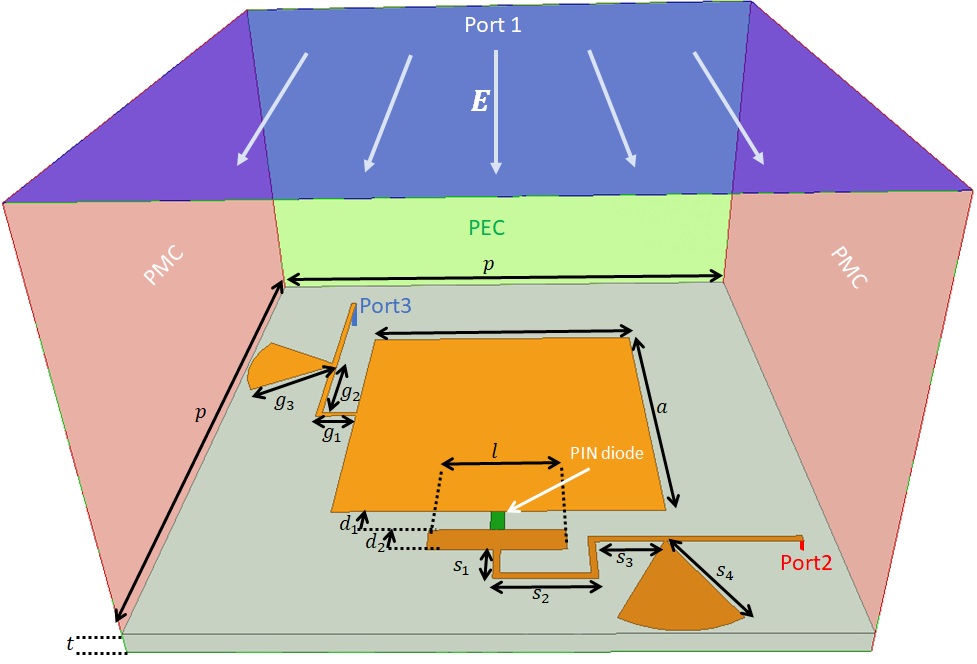}
\caption{Setup to simulate the constitutive element of the RIS in Ansys HFSS. The designed parameters are listed in Table. 1.}
\label{fig:element}
\end{figure}

The constitutive element of the RIS is a PIN-diode loaded resonant patch element implemented on a grounded substrate. This element and its design parameters are shown in Fig. \ref{fig:element}. This geometry is selected due to its successful performance in previous works \cite{trichopoulos2022design}. Other configurations, such as mushroom structures \cite{sleasman2016microwave}, can also be utilized. We used Ansys HFSS full-wave electromagnetic solver to fine-tune the element's geometrical parameters to exhibit a tunable resonance at around $6$ GHz. This frequency is selected for ease of fabrication, component availability, and simple measurement requirements. The proposed RIS can be easily extended to other frequencies by changing the elements' geometry and/or the switchable components.

The proposed patch element is implemented on a 32 mil-thick grounded Rogers 4003 substrate (dielectric constant of 3.55 and loss tangent of 0.0027). The switchable component is selected to be a PIN diode manufactured by Infineon Technologies with part number BAR5002VH6327XTSA1. This diode is modeled as a $0.15$pF capacitor when unbiased or \textit{off} and as a $2$ ohm resistor when it is conducting or \textit{on}. Using simulation in Ansys HFSS in a setup shown in Fig. \ref{fig:element}, the element's geometrical parameters are designed to ensure a resonance frequency around 6 GHz. In this setup, the element is assumed to be placed in an infinite periodic array (implemented by PEC and PMC boundary conditions in the setup of Fig. \ref{fig:element}). For simplicity, the element spacing, $p$, is selected to be around half a wavelength. Closer element spacing is possible, but does not add more degrees of freedom since the wave variation inside the cavity is on the order of half a wavelength. Hence, there is an important distinction between the design of the RIS for this application compared to the ones for beamforming: while beamforming RISs can benefit from smaller element spacing to obtain more control over reflected patterns (especially when using a single bit of modulation), decreasing element spacing here would only increase the complexity and cost of the RIS without adding new degrees of freedom. The patch size is around half the guided wavelength inside the substrate. The element orientation and the placement of the diode are such that the diode can only alter the reflection coefficient of the incident wave for single polarization (denoted in Fig. \ref{fig:CIR}). To arrive at a design that closely resembles practical implementation, we have also included the ground and DC signal paths (required to bias the PIN diode). To decouple the DC and microwave signals, we have included two radial stubs. Their size is selected to be slightly less than the desired quarter wavelength to allow for the simple routing of the DC and ground signal on the RIS board. The microstrip lines used to route the DC signals are $0.5$ mm wide.

\begin{figure}[ht]
\centering
\includegraphics[width=0.85\linewidth]{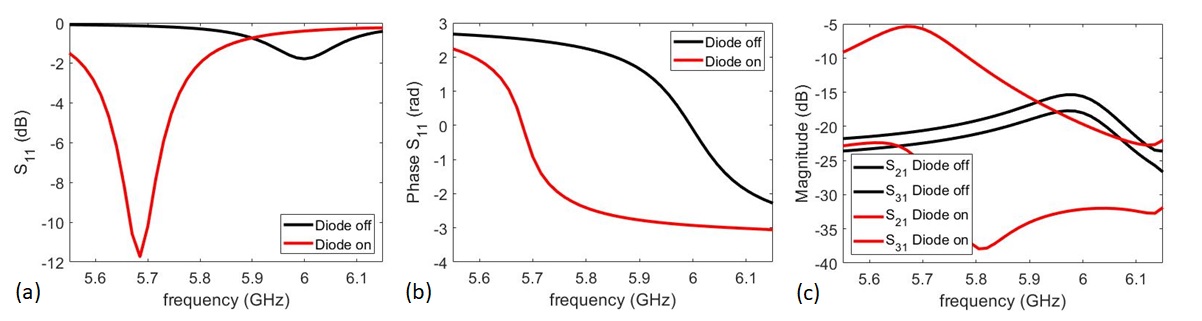}
\caption{(a) Reflection amplitude from the designed element for two states of the diode. (b) Reflection phase of the designed element for two states of the diode. (c) Amplitude of signal coupled to DC biasing circuitry for the two states of the diode.}
\label{fig:elementSparam}
\end{figure}

Using the setup shown in Fig. \ref{fig:element} in Ansys HFSS, we designed the element geometrical parameters with the following goals: tunable resonance near 6 GHz and small leakage of RF signal to the DC biasing lines. The performance of the designed element is shown in Fig. \ref{fig:elementSparam} and the corresponding geometrical parameters are listed in Table \ref{tab:example}. We clearly see a resonance dip in Fig. \ref{fig:elementSparam} (a) for the reflection coefficient around 6 GHz when the diode is \textit{off}. This resonance is shifted to lower frequencies when the diode is biased. It is expected that when the diode is conducting, the effective length of the patch is longer, resulting in a resonance at lower frequencies. The phase difference between the two states of the diode is shown in Fig. \ref{fig:elementSparam} (b). We clearly see a close to $180^\circ$ phase difference between the two states. In other words, the element's effective response changes between PEC and PMC boundary conditions as the diode is turned \textit{off} and \textit{on}. This performance satisfies our first requirement for the RIS elements: the ability to change the boundary condition of the computer chassis (cavity). Examining results in Fig. \ref{fig:elementSparam} (c), we can also confirm the RF and DC signal have been decoupled by using the radial stubs, as the signal observed at the of port 2 and port 3 of the setup in Fig. \ref{fig:element} are well below -10 dB around the design frequency (6 GHz). In this work, we opted for using a structure with DC biasing lines implemented on the same layer as the reflective patches. Such a choice limits the size of the radial stubs and can also result in slight interaction with the incident signal. In applications focused on beamforming using reconfigurable reflective metasurfaces, the biasing circuitry is usually implemented in another layer, preventing the issues listed above at the cost of more complicated and expensive fabrication. For the application considered here, which does not require beamforming, the impact of the DC bias lines can be tolerated since the role of the metasurface is distinct: it needs to change the boundary condition of a disordered cavity. Small changes due to the DC bias line do not interfere with this role. It in fact, may add another layer of disorder.

The fabricated structure for the proof of concept here consisted of only 16 elements. A larger structure with more elements can definitely improve the overall performance, as shown in previous works on wavefront shaping in disordered cavities \cite{del2016spatiotemporal}. However, computer chassis have a limited space, and adding a very large RIS, as is customary in beamforming applications, will not be ideal. Furthermore, the proposed operation does not require a very large aperture since every element can add a new degree of freedom. This is better understood by comparing the case of an RIS with 16 elements, which offers 65536 boundary condition states, and an RIS with 17 elements, which offers 131072 states. The choice of 16 elements was made based on a trade-off to keep fabrication and cost complexity at a minimum for this proof of concept demonstration.  
\begin{table}[ht]
\centering
\begin{tabular}{|l|l|l|l|l|l|l|l|l|l|l|l|l|l|l|l|l|l|}
\hline
Parameter& $p$ &$a$  & $l$ & $t$ &$d_1$& $d_2$& $g_1$ & $g_2$ & $g_3$ & $s_1$ & $s_2$& $s_3$& $s_4$\\
\hline
Value(mm)& 24  &12.5 & 5 & 0.812 & 1   & 1    & 1.75  & 4     &  4.1  & 1.375 & 3.625 & 2.46&  4.1\\
\hline
\end{tabular}
\caption{\label{tab:example} The designed parameters for the constitutive element of the RIS shown in Fig. \ref{fig:element}.}
\end{table}

\subsection*{Measurement Setup}
For our experimental proof of concept, we utilized the chassis of a typical Dell Precision Tower 5810. This computer is non-functional because the motherboard has been removed. Adding a motherboard would only increase the multiple scattering within the chassis, making the need for the proposed device to mitigate multiple scattering even more critical. To add the RIS inside the chassis, we removed one metallic wall and replaced it with a Styrofoam panel that was covered with copper tape (side facing the inside of the chassis). The RIS was then mounted on the copper-taped side of the Styrofoam. Additionally, we made several holes in the Styrofoam to accommodate RF cables for connecting the transmitting and receiving antennas, as well as the DC and ground signals needed for the RIS's operation. For transmitting and receiving signals within this chassis, we used commercially available ultrawideband antennas (specifically, \href{https://www.elecbee.com/en-31285-UWB-Ultra-wideband-Antenna-Working-Frequency-2.4-10.5GHz?utm_term=&utm_campaign=shopping_%E7%BE%8E%E5%9B%BD2021/03/05&utm_source=adwords&utm_medium=ppc&hsa_acc=9958698819&hsa_cam=12473735731&hsa_grp=115457242501&hsa_ad=502747062194&hsa_src=g&hsa_tgt=pla-1186173532455&hsa_kw=&hsa_mt=&hsa_net=adwords&hsa_ver=3&gclid=CjwKCAiAjs2bBhACEiwALTBWZZLN3255AiAAsZr7efztauQMvoqMSGAtCHdsojc8S2xecgdEFj98dxoC-okQAvD_BwE}{Elecbee} antennas were used). This choice was in part to accommodate large bandwidth when measuring the signal between the transmitter and receiver antennas, which were connected to a ZNB43 network analyzer. We conducted our measurement without calibration of the network analyzer to save time. It is worth emphasizing that we have conducted around 65536 measurements for each case. Calibrating each measurement could substantially increase the measurement time. Furthermore, since the cables are low loss, the lack of calibration only introduced small amount of decay in the measurement signal and extra delay in the time domain (that is why the main peak in the CIR shown in Fig. \ref{fig:CIR} appear with a delay that may seem extensive for a chassis environment). The RIS was controlled using an Arduino through a network of daisy-chained shift registers. Both measurement and Arduino were controlled via a desktop computer using MATLAB codes.

\section*{Data availability}

Data that supports the findings of this study is available from the corresponding author (M. I.) upon reasonable request.

\bibliography{sample}

%\noindent LaTeX formats citations and references automatically using the bibliography records in your .bib file, which you can edit via the project menu. Use the cite command for an inline citation, e.g.  \cite{Hao:gidmaps:2014}.

%For data citations of datasets uploaded to e.g. \emph{figshare}, please use the \verb|howpublished| option in the bib entry to specify the platform and the link, as in the \verb|Hao:gidmaps:2014| example in the sample bibliography file.

%\section*{Acknowledgments}

%The authors would like to thank Niccolo Lemonis for his help with data processing.

\section*{Author contributions statement}

M.I. proposed the concept. A.C. and R.R. designed the RIS, L.M., J.S. and J.V. designed the Arduino/breadboard circuit, J.S. and L.M. prepared the software to operate the circuit. All authors constructed the measurement setup and reviewed the manuscript. This work was done while all authors were at Arizona State University.

\section*{Competing Interests}
The authors declare no competing interests.

\end{document}